\documentclass[aps,prd,superscriptaddress,notitlepage,showpacs]{revtex4-1}

\usepackage{graphicx}
\usepackage{amsmath}
\usepackage{amssymb}
\usepackage{color}
\usepackage{bm}

\definecolor{purple}{rgb}{0.8,0,0.6}

\newcommand{\TITLE}{Chiral asymmetry in QED matter in a magnetic field}

\begin{document}

\title{\TITLE}

\date{\today}

\preprint{UWO-TH-13/12}

\author{E. V. Gorbar}
%\email{gorbar@bitp.kiev.ua}
\affiliation{Department of Physics, Taras Shevchenko National Kiev University, Kiev, 03022, Ukraine}
\affiliation{Bogolyubov Institute for Theoretical Physics, Kiev, 03680, Ukraine}

\author{V. A. Miransky}
%\email{vmiransk@uwo.ca}
\affiliation{Department of Applied Mathematics, Western University, London, Ontario N6A 5B7, Canada}

\author{I. A. Shovkovy}
%\email{igor.shovkovy@asu.edu}
\affiliation{School of Letters and Sciences, Arizona State University, Mesa, Arizona 85212, USA}

\author{Xinyang Wang}
%\email{xwang176@asu.edu}
\affiliation{Department of Physics, Arizona State University, Tempe, Arizona 85287, USA}

\begin{abstract}
We calculate the electron self-energy in a magnetized QED plasma to the leading perturbative order 
in the coupling constant and to the linear order in an external magnetic field. We find that the chiral asymmetry of 
the normal ground state of the system is characterized by two new Dirac structures. One of them 
is the familiar chiral shift previously discussed in the Nambu--Jona-Lasinio model.
The other structure is new. It formally looks like that of the chiral chemical potential but is an odd 
function of the longitudinal component of the momentum, directed along the magnetic field. The 
origin of this new parity-even chiral structure is directly connected with the long-range character 
of the QED interaction. The form of the Fermi surface in the weak magnetic field is determined.
\end{abstract}


\maketitle

\section{Introduction}

The study of chiral asymmetry in the normal state of relativistic matter in a magnetic field is at present 
a very active research area in particle physics, with important developments also in condensed matter
physics. Although a chirally asymmetric response of relativistic matter to an external magnetic field 
was discovered long ago \cite{Vilenkin,Zhitnitsky}, the recent interest in this subject is connected with 
the theoretical prediction of the chiral magnetic effect in 
Refs.~\cite{Kharzeev:2004ey,Kharzeev:2007tn,Kharzeev:2007jp,Fukushima:2008xe} and the
subsequent experimental observation of the STAR Collaboration of charge separation in heavy 
ion collisions \cite{collisions,Selyuzhenkov:2011xq,Wang:2012qs,Ke:2012qb,Adamczyk:2013kcb}.

The physical and mathematical reasons for the chiral asymmetry in relativistic matter in a magnetic 
field are quite transparent (for a recent elegant exposition, see Ref.~\cite{Basar-Dunne}). In a free 
theory, the magnetic field $\mathbf{B}$ projects the spins of fermions on the lowest Landau level 
(LLL) along the direction of the magnetic field. Since fermions can freely propagate in a magnetic 
field only along or opposite to the direction of $\mathbf{B}$ and the helicity of massless fermions is 
the same as their chirality, magnetized relativistic matter responds chirally asymmetrically to the magnetic 
field. This leads to the appearance of a nondissipative axial current $\mathbf{j}_5=e\mathbf{B}
\mu/(2\pi^2)$ \cite{Vilenkin,Zhitnitsky}. This effect is known as the chiral separation effect in the literature 
(see, e.g., Sec. 2 in Ref.~\cite{Fukushima:2012vr}).

It was suggested in Refs.~\cite{Kharzeev:2004ey,Kharzeev:2007jp} that topological charge 
changing transitions in QCD during heavy ion collisions may result in the appearance of 
metastable domains with ${\cal P}$ and ${\cal CP}$ breaking with chirality induced in quark-gluon
plasma by the axial anomaly. Phenomenologically, to mimic the effect of topological charge 
changing transitions, it was proposed in \cite{Fukushima:2008xe} to introduce a chiral 
chemical potential $\mu_5$. This chemical potential couples to the difference between 
the number of left- and right-handed fermions and enters the Lagrangian density through 
the term $\mu_5\bar{\psi}\gamma^0\gamma^5\psi$. This produces a chiral asymmetry and 
leads to a nondissipative electric current $\mathbf{j}=e^2\mathbf{B}\mu_5/(2\pi^2)$ in the presence 
of an external magnetic field.

First studies of interaction effects on the chiral asymmetry of relativistic matter in a magnetic field were 
done in Refs.~\cite{chiral-shift-1,chiral-shift-2,Fukushima:2010zza} by using Nambu--Jona-Lasinio (NJL)-like 
models with local interaction. In particular, by using the Schwinger--Dyson (gap) equation, it was found 
that the interaction unavoidably generates a chiral shift parameter $\Delta$ \cite{chiral-shift-1,chiral-shift-2} 
when the fermion density is nonzero. This parameter enters the low-energy effective Lagrangian density 
through the term $\Delta\bar{\psi}\gamma^3\gamma^5\psi$ and determines a relative shift of the momenta 
in the dispersion relations for opposite chirality fermions $k^3 \to k^3 \pm\Delta$, where $k^3$ is the 
longitudinal component of the momentum, directed along the magnetic field. Furthermore, as shown in 
Refs.~\cite{chiral-shift-1,chiral-shift-2,chiral-shift-3}, the chiral shift $\Delta$ is responsible for an additional 
contribution to the axial current. (Note, however, that the dynamical generation of the chiral shift cannot 
and does not modify the form of the chiral anomaly relation \cite{chiral-shift-3}.)

The chiral asymmetry for noninteracting fermions exists only in the LLL. On the other hand,  
the chiral shift parameter found in Refs.~\cite{chiral-shift-1,chiral-shift-2} in the NJL model is 
the same for all Landau levels. This means that the whole Fermi surfaces of the left- and 
right-handed fermions are shifted relative to each other by $2\Delta$ in the longitudinal 
direction in momentum space. Such an unusual ground state closely resembles that in 
Weyl semimetals in condensed matter physics \cite{Wan,Balents}, in which quasiparticles 
are described by the Weyl equation.

We should also mention that the absence of the chiral shift is not protected by any symmetry. 
The chiral shift breaks time reversal $\cal{T}$ and the rotational symmetry $SO(3)$ to the $SO(2)$ 
symmetry of rotations about the axis set by the magnetic field. Since these symmetries are 
explicitly broken in magnetized relativistic matter, it is natural to expect that the generation of 
chiral shift and chiral asymmetry in all Landau levels is not a special property of the NJL model, but
a model-independent phenomenon.

In this connection, it is instructive to discuss the equation 
for $\Delta$ in the perturbation theory in the NJL model \cite{chiral-shift-2}. In the first order in the 
coupling constant $G_{\rm int}$, the value of $\Delta$ is proportional to the axial current in the free 
theory \cite{chiral-shift-2}, $\Delta= -\frac{1}{2}G_{\rm int} \langle j^{3}_5\rangle_0$, which 
is completely defined by the topological contribution $\langle j^{3}_5\rangle_0$ from the LLL 
\cite{Zhitnitsky}. Therefore, the interaction causes the chiral asymmetry of the LLL 
to propagate to higher Landau levels including those around the whole Fermi surface.

It was found in Refs.~\cite{chiral-shift-1,chiral-shift-2,chiral-shift-3} that the interactions in the 
NJL model not only induce chiral asymmetry in the higher Landau levels but also lead to 
the dynamical contribution of the higher levels to the axial current. In the recent paper
\cite{Gorbar:2013upa}, the leading radiative corrections to the axial current were calculated 
in dense QED in a magnetic field. It was found that, like in the NJL model, the axial current 
is not fixed by the chiral anomaly relation and thus does not coincide with the expression
in the free theory, solely provided by the LLL. Instead, it receives nontrivial radiative 
corrections produced at all Landau levels. Because of that, it is natural to expect that, 
like in the NJL model, the chiral asymmetry in QED will be induced by interactions in 
higher Landau levels, too.

This point was the main motivation for the present paper, in which we analyze the chiral asymmetry
in the first order of perturbation theory in the QED coupling $\alpha=e^2/(4\pi)$. The paper is 
organized as follows. In Sec.~\ref{Sec:Model}, we introduce the model and set up the notation. 
The calculation of the self-energy is presented in Sec.~\ref{Sec:self-energy}. Its projection onto 
Landau levels is considered in Sec.~\ref{Sec:Landau-level-expansion}. The self-energy in the 
weak magnetic field limit is studied in Sec.~\ref{Sec:Self-energy-pseudomomentum}.
The modified dispersion relations for the left- and right-handed fermions with the inclusion of the 
chiral asymmetry are calculated in Sec.~\ref{Sec:Asymmetry}. Finally, our discussion of the main 
results is given in Sec.~\ref{Sec:Conclusion}. We use the units with $\hbar=c=1$.

\section{Model}
\label{Sec:Model}

The Lagrangian density of QED in an external magnetic field is given by
\begin{equation}
{\cal L}=-\frac{1}{4}F^{\mu\nu}F_{\mu\nu}+\bar{\psi}\left( i\gamma^{\nu}{\cal D}_{\nu}+\mu
\gamma^0-m\right)\psi,
\label{Lagrangian}
\end{equation}
where the covariant derivative is ${\cal D}_{\nu} =\partial_{\nu}-i e A^{\rm ext}_{\nu}-i e A_{\nu}$ and $m$ is the bare
fermion mass (here we use the notation of Ref.~\cite{Peskin}, but with the opposite sign of the electric charge, $e\to -e$). 
In order to describe a nonzero density plasma, we introduced the fermion chemical potential $\mu$.
Without loss of generality, we assume that the external magnetic field $\mathbf{B}$ points 
in the $z$ direction. The components of the conventional vectors, including those of the 
vector potential $\mathbf{A}^{\rm ext}$, are identified with the {\em contravariant} components. 
(Note that the components of the spatial gradient $\bm{\nabla}$ are given by $\partial_k\equiv -\partial^k $.) 
In the rest of this paper, we use the vector potential in the Landau gauge, $\mathbf{A}^{\rm ext}= (0, x^1 B,0)$.

Before proceeding to the main part of the analysis, we should recall that the
fermion propagator in the presence of an external magnetic field is not a translation invariant 
function. From the physics viewpoint, this reflects the fact that charged fermions cannot have well 
defined momenta in the spatial directions perpendicular to the direction of the magnetic field. 
In the case of a uniform magnetic field, however, the propagator can be represented as a product 
of the Schwinger phase \cite{Schwinger:1951nm}, which is the only part that spoils the translational invariance, and 
a translationally invariant function, i.e.,
\begin{equation}
S(x,y)=e^{i\Phi(x,y)}\bar{S}(x-y).
\label{fermion-propagator}
\end{equation}
In the Landau gauge used here, the Schwinger phase is given
by $\Phi(x,y)=-eB(x_1+y_1)(x_2-y_2)/2$. The Fourier transform of the translation 
invariant part of the fermion propagator $\bar{S}(x-y)$ reads \cite{Chodos:1990vv}
\begin{equation}
\bar{S}(k) = 2 i  e^{-k_{\perp}^2 \ell^2 }  \sum_{n=0}^{\infty} \frac{(-1)^n 
D_n(k)}{[k_0+\mu+i\epsilon\,\mbox{sgn}(k_0)]^2-2n|eB| -(k^3)^2-m^2},
\label{prop-momentum}
\end{equation}
where $\ell\equiv 1/\sqrt{|eB|}$ is the magnetic length and 
\begin{eqnarray}
D_n(k) &=& \left[\gamma^0 (k_0+\mu) -\gamma^3 k^3 +m\right]
\left[ L_{n}\left(2k_{\perp}^2 \ell^2\right) {\cal P}_{+} 
-  L_{n-1}\left(2k_{\perp}^2 \ell^2\right){\cal P}_{-}\right] 
 +2(\bm{\gamma}_{\perp}\cdot\mathbf{k}_\perp) L^{1}_{n-1}\left(2k_{\perp}^2 \ell^2\right),
\label{prop-D-n-momentum}
\end{eqnarray}
where ${\cal P}_{\pm}=\left[1\pm i\,\mbox{sgn}(eB)\gamma^1\gamma^2\right]/2$ are spin projectors 
and $L^{(\alpha)}_n(x)$ are generalized Laguerre polynomials \cite{Gradstein_Ryzhik}. 
(For the proper-time Schwinger representation of the propagator generalized to the case of finite 
density, see Appendix~A in Ref.~\cite{Gorbar:2013upa}.) Note that, despite the appearance, this 
is not a conventional momentum-space representation of the fermion propagator. Strictly 
speaking, after the Schwinger phase is removed, the Fourier transform of $\bar{S}(x-y)$ 
cannot even be interpreted as an actual propagator. The reason is that from the physics viewpoint the 
transverse components $\mathbf{k}_{\perp}$ of four-vector $k$ are not good quantum numbers 
for classifying fermionic states in a magnetic field. To avoid a potential confusion, therefore, we 
call $k$ a pseudomomentum in what follows. It is instructive to mention though that, in the limit 
of large pseudomomentum or weak magnetic field (i.e., $\mathbf{k}^2_{\perp} \gg |eB|$), the effects 
of the Schwinger phase can be neglected and pseudomomentum can be interpreted as 
an approximate (or ``quasiclassical") fermion's momentum. We will utilize this 
fact in our weak field analysis in Sec.~\ref{Sec:Asymmetry}.

In the analysis below, we will be using the photon propagator in the Feynman gauge. 
In momentum space, its explicit form reads
\begin{equation}
D_{\mu\nu}(q)=-i\frac{g_{\mu\nu}}{q^2_{\Lambda}} \equiv  
-i\left(\frac{g_{\mu\nu}}{q_0^2-\mathbf{q}^2-m_\gamma^2+i\epsilon}
-\frac{g_{\mu\nu}}{q_0^2-\mathbf{q}^2-\Lambda^2+i\epsilon} \right),
\label{photon-propagator}
\end{equation}
where, following the same methodology as in Ref.~\cite{Gorbar:2013upa}, we introduced a nonzero 
photon mass $m_{\gamma}$ (since unlike the radiative corrections calculated in \cite{Gorbar:2013upa} the fermion self-energy is
regular in the infrared region, we will set $m_{\gamma}=0$ in the final results) and an ultraviolet cutoff $\Lambda$ that will serve as an
infrared and ultraviolet regulators, respectively, at the intermediate stages of calculations.

\section{Fermion self-energy in a magnetic field}
\label{Sec:self-energy}

To leading order in coupling constant $\alpha=e^2/(4\pi)$, the fermion self-energy in QED is 
given by 
\begin{equation}
\Sigma(x,y)=-4i\pi\alpha\gamma^\mu S(x,y) \gamma^\nu D_{\mu\nu}(x-y).
\label{self-energy}
\end{equation}
Notice that the self-energy $\Sigma(x,y)$ is not a translation invariant function when the external 
magnetic field is present. By making use of the propagator $S(x,y)$ in the Schwinger representation 
(\ref{fermion-propagator}), it is convenient to factor out the Schwinger phase in the self-energy, i.e., 
$\Sigma(x,y)=\exp\left[i\Phi(x,y)\right]\bar{\Sigma}(x-y)$. The Fourier transform (or, equivalently, 
pseudomomentum representation) of the translation invariant function $\bar{\Sigma}(x-y)$ is 
given by the following expression:
\begin{equation}
\bar{\Sigma} (p) = - 4i\pi\alpha \int\frac{ d^4 k}{(2\pi)^4}
\gamma^{\mu}\,\bar{S} (k)\gamma^{\nu} D_{\mu\nu}(k-p)=-4\pi\alpha \int\frac{ d^4 k}{(2\pi)^4}
\gamma^{\mu}\,\bar{S} (k)\gamma_{\mu}\frac{1}{(k-p)^2_{\Lambda}},
\label{self-energy-momentum}
\end{equation}
where $\bar{S}(k)$ is the Fourier transform of the translation invariant part of the fermion propagator, 
given in Eq.~(\ref{prop-momentum}), and $D_{\mu\nu}(q)$ is the photon propagator in momentum 
space, given in Eq.~(\ref{photon-propagator}). By taking into account the Dirac structure of $\bar{S}(k)$ 
and the identity $\gamma^{\mu}\gamma^{\nu}\gamma_{\mu}=-2\gamma^{\nu}$, it is straightforward 
to show that the resulting pseudomomentum representation of the self-energy (\ref{self-energy-momentum}) 
has the following Dirac structures:
\begin{eqnarray}
\bar{\Sigma} (p) &=& 
- \gamma^0{\delta\mu(p)} 
+ p^3\gamma^3\,{\delta v_{3}(p)} 
+ (\bm{\gamma}_{\perp}\cdot\mathbf{p}_{\perp}){\delta v_{\perp}(p)} 
+ {\cal M}(p) 
- i \gamma^1 \gamma^2 \tilde{\mu}(p)
- \gamma^3\gamma^5 \Delta(p)
- \gamma^0\gamma^5 \mu_5(p).
\label{self-energy-general}
\end{eqnarray}
The first four Dirac structures in Eq.~(\ref{self-energy-general}) are standard and are present 
in the fermion self-energy also when the magnetic field is absent. The functions ${\delta\mu(p)}$, 
${\delta v_{3}(p)}$, and $\delta v_{\perp}(p)$ define the wave-function renormalization and the 
modification of the (longitudinal and transverse) fermion velocities. Note that in the absence of 
a magnetic field $\delta v_3=\delta v_{\perp}$. 
The contribution with the unit matrix ${\cal M} (p)$ gives a correction to the fermion mass function. 
As for the last three terms in the self-energy (\ref{self-energy-general}), they are obtained from 
Eq.~(\ref{prop-D-n-momentum}) by taking the terms with the $i\gamma^1\gamma^2$ 
matrix in the spin projectors ${\cal{P}}_{\pm}$. Obviously, these Dirac structures are present only if there 
is a magnetic field. The terms with $\tilde{\mu} (p)$ and $\Delta(p)$ are the anomalous 
magnetic moment function and chiral shift, respectively. They are generated \cite{chiral-shift-1,chiral-shift-2} 
in the NJL model, too, where they are constants. Here, due to the long-range character of the QED interaction, however, these
functions generally depend on the pseudomomentum.

The last term in the self-energy (\ref{self-energy-general}) presents a qualitatively new type of contribution in QED.
As we show below, it has the form $\mu_5(p)\equiv p_3 f(p)$, 
where $f(p)$ is a dimensionless function. This new contribution comes as a result of the long-range QED interaction 
and, thus, has no analog in the NJL model. If $\mu_5(p)$ were a constant and did not depend on pseudomomentum, 
it would be identical with the chiral chemical potential $\mu_5$ \cite{Fukushima:2008xe,Kharzeev:2009fn}
and would, therefore, break parity. Considering that neither the external magnetic field nor the electromagnetic 
interaction breaks parity, the genuine chiral chemical potential cannot 
be generated in perturbation theory. Instead, we find that the function $\mu_5(p)$ in the self-energy 
(\ref{self-energy-general}) is an odd function of the $p^3$-component of momentum and, 
therefore, is even under parity.

Combining Eqs.~(\ref{self-energy-momentum}) and (\ref{self-energy-general}), one can determine the coefficient functions
$\delta\mu(p)$, $\delta v_3(p)$, etc., of the translation invariant part of the fermion self-energy in pseudomomentum space. 
However, as we mentioned above, the transverse components $\mathbf{k}_{\perp}$ of four-vector $k$ are not good 
quantum numbers for fermions in a magnetic field. Therefore, although the pseudomomentum representation in 
Eq.~(\ref{self-energy-general}) will be extremely useful in the limit of a weak magnetic field and we will use it 
to analyze the chiral asymmetry of QED matter in Sec.~\ref{Sec:Asymmetry}, it is still highly desirable to
obtain the self-energy in the much more natural Landau level representation, in which all dynamical functions 
of the self-energy are projected onto specific Landau levels. Clearly, the dynamical functions projected onto 
specific Landau levels provide a direct physical interpretation and are essential for determining the 
interaction-induced modifications of the dispersion relations of quasiparticles in each individual Landau level $n$. 
In the next section, we show how such a projection is realized and provide explicit formulas expressing the
dynamical functions $\delta\mu_n$, $\delta v_{3,n}$, etc., projected onto specific Landau levels through 
the self-energy (\ref{self-energy-momentum}).

\section{Projection onto Landau levels}
\label{Sec:Landau-level-expansion}

One of the standard approaches to treat quantum field theory of charged particles in 
an external magnetic field makes use of the Ritus eigenfunctions \cite{Ritus-1,Ritus-2}. 
In this study, however, we advocate a different approach that was recently developed in 
Ref.~\cite{Gorbar:2011kc} in a study of graphene in a magnetic field. From the technical 
viewpoint, the key difference between the two methods lies in the use of the complete sets of 
eigenfunctions of different operators. In the Ritus method, one uses the eigenfunctions of the operator 
$(\bm{\pi}\cdot\bm{\gamma})$ with a nontrivial Dirac structure and, thus, treats both the 
orbital and spinor parts of the fermion kinematics in a uniform fashion. In the method of 
Ref.~\cite{Gorbar:2011kc}, on the other hand, the eigenfunctions of the scalar operator 
$\bm{\pi}^2$ are used. This operator includes only the orbital part of the kinematics and, 
thus, requires one to treat the spin part separately. The seeming inconvenience of dealing with 
the spinor part separately in the second method, in fact, appears to offer many advantages, 
ranging from a much more transparent interpretation of various Dirac structures 
in the propagator and self-energy to significant technical simplifications in calculations.
  
In the rest of this section, we give a detailed derivation of the expansion of
the fermion self-energy over Landau levels. We will start by writing down the self-energy
with all the Dirac structures in Eq.~(\ref{self-energy-general}) in the coordinate space 
(clearly, the Fourier transformation of the self-energy in the pseudomomentum representation 
and its multiplication by the Schwinger phase do not change the Dirac structure of the self-energy). 
Thus, we have
\begin{eqnarray}
\Sigma(x,y) &=&
\left[-\gamma^0 {\delta\mu} 
+\pi^3\gamma^3  \delta v_{3}
+(\bm{\pi}_{\perp}\cdot\bm{\gamma}_{\perp}){\delta v_{\perp}} 
+{\cal M} 
- i\gamma^1\gamma^2\tilde{\mu}
-\gamma^3\gamma^5\Delta
-\gamma^0\gamma^5\mu_{5} \right] \delta^{4}(x-y),
\label{self-energy-coordinate-space}
\end{eqnarray}
where $\mathbf{\pi}_{\perp}$ is the canonical transverse momentum operator, which 
includes the vector potential. Here, $\delta\mu$, $\delta v_{3}$, $\delta v_{\perp}$, ${\cal M}$, 
$\tilde{\mu}$, $\Delta$, and $\mu_5$, are functions of the operators $-i\partial_0$ and $\pi^3$, 
as well as the operator $(\bm{\pi}_{\perp}\cdot\bm{\gamma}_{\perp})^2\ell^2$.

The eigenvalues of the operator $\bm{\pi}_{\perp}^2\ell^2$ are positive odd integers,
$2N+1$, where $N=0,1,2,\ldots$ is the orbital quantum number \cite{Landau}. The corresponding 
eigenfunctions $\psi_{Np}(\mathbf{r}_{\perp})$ are well known and have the following explicit form: 
\begin{equation}
\psi_{Np}(\mathbf{r}_{\perp}) =\frac{1}{\sqrt{2\pi \ell}}\frac{1}{\sqrt{2^NN!\sqrt{\pi}}}
H_N\left(\frac{x}{\ell}+p\ell\right)e^{-\frac{1}{2\ell^2}(x+p\ell^2)^2} e^{i  py\, \mbox{\scriptsize sgn}(eB)},
\end{equation}
where $H_n(x)$ are Hermite polynomials \cite{Gradstein_Ryzhik}.
By making use of the completeness of these eigenfunctions 
\begin{eqnarray}
\delta^2(\mathbf{r}_{\perp}-\mathbf{r}^{\prime}_{\perp})=
\sum_{N=0}^{\infty}\int^{+\infty}_{-\infty} dp\,\,\psi_{Np}(\mathbf{r}_{\perp})\psi^*_{Np}(\mathbf{r}^{\prime}_{\perp}),
\label{completeness}
\end{eqnarray}
one can rewrite the self-energy (\ref{self-energy-coordinate-space}) in the form
\begin{eqnarray}
\Sigma(x,y)&=&\sum_{N=0}^{\infty}\sum_{s=\pm}\int\frac{dp_0 dp^3 dp}{(2\pi)^2}
e^{-ip_0(x_0-y_0)+ip^3(x^3-y^3)}
\nonumber\\
&&\times\left[-\gamma^0\delta\mu+p^3\gamma^3\delta v_3
+(\bm{\pi}_{\perp}\cdot\bm{\gamma}_{\perp}){\delta v_{\perp}} 
+{\cal M}
- i\gamma^1\gamma^2\tilde{\mu}-\gamma^3\gamma^5\Delta-\gamma^0\gamma^5\mu_{5}
\right] \,{\cal P}_s\,\psi_{Np}(\mathbf{r}_{\perp}) \psi_{Np}^{*}(\mathbf{r}_{\perp}^\prime),
\label{self-energy-coordinate-levels}
\end{eqnarray}
where $\delta\mu$, $\delta v_3$, $\ldots$, $\mu_5$ are now functions of $p_0$, $p^3$, and the operator 
$(\bm{\pi}_{\perp}\cdot\bm{\gamma}_{\perp})^2\ell^2$.  In Eq.~(\ref{self-energy-coordinate-levels}), we 
also inserted the unit matrix in the form of the sum of spin projectors, i.e., $1=\sum_{s=\pm}{\cal P}_s$. 
It is easy to see that any function of $(\bm{\pi}_{\perp}\cdot\bm{\gamma}_{\perp})^2\ell^2$ acting on 
${\cal P}_s\,\psi_{Np}$ reduces to a constant in the $n$th Landau level. This is a consequence of 
the following identity:
\begin{equation}
(\bm{\pi}_{\perp}\cdot\bm{\gamma}_{\perp})^2\ell^2\,{\cal P}_s\,\psi_{Np}
=-(\bm{\pi}^2_{\perp}+ieB\gamma^1\gamma^2)\ell^2\,{\cal P}_s\,\psi_{Np}
=-2 n\,{\cal P}_s\,\psi_{Np},
\end{equation}
where $n\equiv N+(1+s)/2$ is the standard Landau level quantum number. This allows us to 
rewrite Eq.~(\ref{self-energy-coordinate-levels}) as follows:
\begin{eqnarray}
\Sigma(x,y)&=&\sum_{N=0}^{\infty}\sum_{s=\pm}\int\frac{dp_0 dp^3 dp}{(2\pi)^2}
e^{-ip_0(x_0-y_0)+ip^3(x^3-y^3)} \Big[-\gamma^0\delta\mu_n+p^3\gamma^3\delta v_{3,n}
+(\bm{\pi}_{\perp}\cdot\bm{\gamma}_{\perp}){\delta v_{\perp,n}} 
+{\cal M}_n\nonumber\\
&&
- i\gamma^1\gamma^2\tilde{\mu}_n-\gamma^3\gamma^5\Delta_n-\gamma^0\gamma^5\mu_{5,n}
\Big] \,{\cal P}_s\,\psi_{Np}(\mathbf{r}_{\perp}) \psi_{Np}^{*}(\mathbf{r}_{\perp}^\prime),
\label{SigmaXY_Eigenstates}
\end{eqnarray}
where the coefficient functions $\delta\mu_n$, $\delta v_{3,n}$, etc., depend on energy $p_0$ 
and longitudinal momentum $p^3$. Now, by taking into account the relation
\begin{equation}
(\bm{\pi}_{\perp}\cdot\bm{\gamma}_{\perp}) \ell \, \psi_{Np}(\mathbf{r}_{\perp}) = 
i \gamma^1  \left[
\sqrt{2(N+1)}{\cal P}_{+} \psi_{N+1,p}(\mathbf{r}_{\perp}) 
-\sqrt{2N}{\cal P}_{-} \psi_{N-1,p}(\mathbf{r}_{\perp}) 
\right]
\label{pi-gamma-relation}
\end{equation}
and using formula 7.377 from Ref.~\cite{Gradstein_Ryzhik},
\begin{equation}
\int\limits_{-\infty}^\infty\,e^{-x^2}H_m(x+y)H_n(x+z)dx
=2^n\pi^{1/2}m!z^{n-m}L_m^{n-m}(-2yz),
\label{HnHm-integral}
\end{equation}
we can perform the integration over $p$ in Eq.~(\ref{SigmaXY_Eigenstates}). As expected, the result takes 
the form of a product of the Schwinger phase and a translationally invariant function, i.e.,
\begin{equation}
\Sigma(x,y) = e^{i\Phi(x,y)} \bar{\Sigma}(x-y) ,
\end{equation}
where the latter is given by 
\begin{eqnarray}
\bar{\Sigma}(x)&=&\frac{e^{-\xi/2}}{2\pi\ell^2}\sum_{n=0}^{\infty}
\int \frac{dp_0 dp^3 }{(2\pi)^2}  e^{-ip_0x_0+ip^3x^3} 
\Big\{
\left(-\gamma^0 \delta\mu_{n} 
+p^3\gamma^3  \delta v_{3,n}
- i\gamma^1\gamma^2\tilde{\mu}_{n}
-\gamma^3\gamma^5\Delta_{n}
-\gamma^0\gamma^5\mu_{5,n}
+{\cal M}_{n} \right)\nonumber\\
&&\times \left[ L_{n}(\xi){\cal P}_{-}+L_{n-1}(\xi){\cal P}_{+}\right]
-\frac{i}{\ell^2}(\mathbf{r}_{\perp}\cdot\bm{\gamma}_{\perp}){\delta v_{\perp,n}} L^{1}_{n-1}(\xi)
\Big\},
\end{eqnarray}
and $\xi=\mathbf{r}^2_{\perp}/(2\ell^2)$. Here $L_{-1}(\xi)=0$ by definition. Performing the 
Fourier transform, we finally find the sought expansion of the fermion self-energy over the 
Landau levels:
\begin{eqnarray}
\bar{\Sigma}(p)&=& 2  e^{-p_{\perp}^2 \ell^2} 
\sum_{n=0}^{\infty}(-1)^n\Big\{
\left(-\gamma^0 \delta\mu_{n} 
+p^3\gamma^3  \delta v_{3,n}
- i\gamma^1\gamma^2\tilde{\mu}_{n}
-\gamma^3\gamma^5\Delta_{n}
-\gamma^0\gamma^5\mu_{5,n}
+{\cal M}_{n} \right) \nonumber\\
&&\times \left[ L_{n}(2p_{\perp}^2 \ell^2){\cal P}_{-} - L_{n-1}(2p_{\perp}^2 \ell^2){\cal P}_{+}\right]
-2 (\bm{\gamma}_\perp\cdot \bm{p}_{\perp}){\delta v_{\perp,n}} L^{1}_{n-1}(2p_{\perp}^2 \ell^2)
\Big\}.
\label{self-energy-LL}
\end{eqnarray} 

In what follows, we will drop the $\delta v_{\perp}$-type corrections to the self-energy. For the 
purposes of this study, this is justified, because such terms neither break the chiral symmetry
nor modify the chiral asymmetry of the ground state. On the other hand, it is necessary to 
keep the terms with ${\delta\mu(p)}$ and ${\delta v_3(p)}$. The reason for this becomes 
obvious after noticing that, when restricted to the subspaces of fixed spin projections, 
${\delta\mu(p)}$ and ${\delta v_3(p)}$ mix up with $\Delta(p)$ and $\mu_5(p)$, respectively. 
The argument can be made explicit by making use of the following identities: 
$\gamma^0 {\cal P}_{\pm} =\pm \mbox{sgn}(eB)\gamma^3\gamma^5 {\cal P}_{\pm}$ and 
$\gamma^3 {\cal P}_{\pm} =\pm \mbox{sgn}(eB)\gamma^0\gamma^5 {\cal P}_{\pm}$. Note 
that a similar argument also necessitates the inclusion of the anomalous magnetic 
moment function $\tilde{\mu}(p)$ whenever the mass function ${\cal M}(p)$ is present.

Equation (\ref{self-energy-LL}) defines the expansion of the fermion self-energy over Landau 
levels. On the other hand, in the leading order of perturbation theory, the self-energy is given 
by Eq.~(\ref{self-energy-momentum}). By combining these equations, it is not difficult to express 
the dynamical functions $\delta\mu_n$, $\delta v_{3,n}$, etc., projected onto specific Landau 
levels, through the self-energy (\ref{self-energy-momentum}). Multiplying these two equivalent 
expressions for the fermion self-energy by 
$\ell^2\pi^{-1}(-1)^{n^\prime}e^{-p^2_{\perp}\ell^2}L_{n^\prime}(2p^2_{\perp}\ell^2){\cal P}_{\pm}$ 
and integrating over the perpendicular momentum $\bm{p}_{\perp}$, we arrive at the following set 
of equations:
\begin{eqnarray}
\left[-\gamma^0 \delta\mu_{n} 
+p^3\gamma^3  \delta v_{3,n}
+{\cal M}_{n}  
+\mbox{sgn}(eB)\left(\tilde{\mu}_{n}
+ \gamma^0\Delta_{n}
+ \gamma^3\mu_{5,n}\right)
\right] {\cal P}_{-} &=& I_n{\cal P}_{-}\,,
\label{equation-spin-1}
\\
\left[ -\gamma^0 \delta\mu_{n} 
+p^3\gamma^3  \delta v_{3,n}
+{\cal M}_{n}  
- \mbox{sgn}(eB)\left(\tilde{\mu}_{n}
+ \gamma^0\Delta_{n}
+ \gamma^3\mu_{5,n}\right)
\right]{\cal P}_{+} &=& I_{n-1}{\cal P}_{+}\,,
\label{equation-spin-2}
\end{eqnarray}
where
\begin{equation}
I_n = -4i (-1)^n \alpha\ell^2 \int\frac{d^2k_{\parallel}d^2k_{\perp}d^2p_{\perp}}{(2\pi)^4}
e^{-p_{\perp}^2\ell^2}\,L_n(2p^2_{\perp}\ell^2)\,
\gamma^{\mu}\,\bar{S}(k)\gamma^{\nu}
D_{\mu\nu}(p-k).
\label{I_n_definition_1}
\end{equation}
When the free fermion propagator in Eq.~(\ref{I_n_definition_1}) is replaced by the full propagator, which itself 
is a function of $\delta\mu_{n}$, $\delta v_{3,n}$, etc., the above set of equations will become an infinite set of 
the Schwinger-Dyson equations for the dynamical functions.

From Eqs.~(\ref{equation-spin-1}) and (\ref{equation-spin-2}), we obtain the following relations which express 
the dynamical functions projected onto specific Landau levels through the self-energy (\ref{self-energy-momentum}):
\begin{eqnarray}
\delta\mu_{n}  &=&  -\frac{1}{4}\mbox{Tr}\left[\gamma^0\left( I_{n}{\cal P}_{-}+ I_{n-1}{\cal P}_{+}\right)\right] ,
\label{eq-n-1} \\
\Delta_{n}    &=& \frac{\mbox{sgn}(eB)}{4}\mbox{Tr}\left[\gamma^0\left( I_{n}{\cal P}_{-}- I_{n-1}{\cal P}_{+}\right)\right]  ,
\label{eq-n-2}\\
{\cal M}_{n}  &=& \frac{1}{4}\mbox{Tr} \left( I_{n}{\cal P}_{-}+ I_{n-1}{\cal P}_{+}\right)  ,
\label{eq-n-3}\\
\tilde{\mu}_{n}   &=&  \frac{\mbox{sgn}(eB)}{4}\mbox{Tr} \left( I_{n}{\cal P}_{-}- I_{n-1}{\cal P}_{+}\right)   ,
\label{eq-n-4}\\
p^3 \delta v_{3,n} &=& -\frac{1}{4}\mbox{Tr}\left[\gamma^3\left( I_{n}{\cal P}_{-}+ I_{n-1}{\cal P}_{+}\right)\right]    ,
\label{eq-n-5}\\
\mu_{5,n} &=& -\frac{\mbox{sgn}(eB)}{4}\mbox{Tr}\left[\gamma^3\left( I_{n}{\cal P}_{-}- I_{n-1}{\cal P}_{+}\right)\right]    .
\label{eq-n-6}
\end{eqnarray}
The special role of LLL ($n=0$) should be noted here. By taking into account that $I_{-1}=0$, 
we find the following relations between the pairs of parameters: 
$\Delta_{0} =- \mbox{sgn}(eB)\delta\mu_{0} $, 
$\tilde{\mu}_{0} =\mbox{sgn}(eB){\cal M}_{0} $, and 
$\mu_{5,0} =\mbox{sgn}(eB)p^3 \delta v_{3,0} $; i.e.,  only half of them remain independent in LLL. 
From the physics viewpoint, this reflects the spin-polarized nature of the lowest energy level. 

The dynamical functions $\Delta_n$ and $\mu_{5,n}$ for $n \ge 1$ define chiral asymmetry 
in higher Landau levels. Therefore, these functions are of the prime interest for us in the
present paper. In terms of the self-energy (\ref{self-energy-momentum}), we can represent 
$I_n$ in Eq.~(\ref{I_n_definition_1}) as follows:
\begin{equation}
I_n =  (-1)^n\frac{\ell^2}{\pi}\int d^2p_{\perp}
e^{-p_{\perp}^2\ell^2}\,L_n(2p^2_{\perp}\ell^2)\,
\bar{\Sigma}(p).
\label{I_n_definition_2}
\end{equation}
Using it, we may rewrite Eqs.~(\ref{eq-n-2}) and (\ref{eq-n-6}) in the following perhaps more transparent form:
\begin{eqnarray}
\Delta_n 
&=&  \frac{(-1)^{n}}{8} \frac{\ell^2}{\pi} \mbox{sgn}(eB) \int d^2p_{\perp}
e^{-p_{\perp}^2\ell^2}\left[L_n(2p^2_{\perp}\ell^2)+L_{n-1}(2p^2_{\perp}\ell^2)\right]
\mbox{Tr}\left[\gamma^0 \bar{\Sigma}(p)\right]\nonumber\\
&& -  \frac{(-1)^{n}}{8} \frac{\ell^2}{\pi}  \int d^2p_{\perp}
e^{-p_{\perp}^2\ell^2}\left[L_n(2p^2_{\perp}\ell^2)-L_{n-1}(2p^2_{\perp}\ell^2)\right]
\mbox{Tr}\left[\gamma^3\gamma^5 \bar{\Sigma}(p)\right],
\label{chiral-shift-n}
\end{eqnarray}
\begin{eqnarray}
\mu_{5,n} 
&=& -\frac{(-1)^{n}}{8} \frac{\ell^2}{\pi} \mbox{sgn}(eB) \int d^2p_{\perp}
e^{-p_{\perp}^2\ell^2}\left[L_n(2p^2_{\perp}\ell^2)+L_{n-1}(2p^2_{\perp}\ell^2)\right]
\mbox{Tr}\left[\gamma^3 \bar{\Sigma}(p)\right]\nonumber\\
&& + \frac{(-1)^{n}}{8} \frac{\ell^2}{\pi}  \int d^2p_{\perp}
e^{-p_{\perp}^2\ell^2}\left[L_n(2p^2_{\perp}\ell^2)-L_{n-1}(2p^2_{\perp}\ell^2)\right]
\mbox{Tr}\left[\gamma^0\gamma^5 \bar{\Sigma}(p)\right],
\label{chemical-potential-n}
\end{eqnarray}
where $\bar{\Sigma}(p)$ is given by Eq.~(\ref{self-energy-momentum}). These expressions will 
in principle allow us to determine the chiral asymmetry for fermions in higher Landau levels. 
In the general case, however, the calculation of these parameters can be done only with the 
help of numerical methods. In fact, we should mention that our initial attempts at such calculations 
suggest that, despite several technical complications (e.g., highly oscillatory integrand, as well as the need 
to numerically regularize the integral and then perform the renormalization), the problem can possibly 
be solved with moderate computational resources. That, however, is beyond the scope of the present paper.

In the rest of this study, we will concentrate on the weak magnetic field limit. In addition 
to providing some simplifications in the analysis, the corresponding approximation is in fact 
sufficient for practically all stellar applications. Indeed, taking into account that 
\begin{equation}
\frac{|eB|}{\mu^2} \simeq 6 \times 10^{-4}\left(\frac{B}{10^{15}~\mbox{G}}\right)
\left(\frac{100~\mbox{MeV}}{\mu}\right)^2,
\end{equation}
we conclude that the magnetic fields can be treated as weak even in the case of magnetars.
\vspace{5mm}

\section{Weak magnetic field limit}
\label{Sec:Self-energy-pseudomomentum}

Since the photon propagator (\ref{photon-propagator}) does not depend on the magnetic field, to find the self-energy 
(\ref{self-energy-momentum}) in a weak magnetic field we should determine the translation invariant part of the free fermion propagator
$\bar{S}(k)$ in a weak magnetic field. To leading order in $B$, it reads \cite{Gorbar:2013upa}
\begin{equation}
\bar{S}(k) = \bar{S}^{(0)}(k) +\bar{S}^{(1)}(k) +\cdots,
\end{equation}
where $\bar{S}^{(0)}$ is the free electron propagator in the absence of magnetic field
and $\bar{S}^{(1)}$ is the linear in the magnetic field part. They are
\begin{eqnarray}
\bar{S}^{(0)}(k) &=& i \frac{(k_0+\mu)\gamma^0- \mathbf{k}\cdot\bm{\gamma}+m}{(k_0+\mu)^2-
\mathbf{k}^2-m^2},
\label{free-term}
\end{eqnarray}
\begin{eqnarray}
\bar{S}^{(1)}(k) &=&  -eB \frac{(k_0+\mu)\gamma^{0}-k^3\gamma^3+m}
{\left[ (k_0+\mu)^2-\mathbf{k}^2 -m^2\right]^2 } \gamma^1\gamma^2.
\label{linear-term}
\end{eqnarray}
We find more convenient in this section instead of $\bar{S}^{(1)}(k)$ given by Eq.~(\ref{linear-term}) 
to use the following equivalent representation \cite{Gorbar:2013upa}:
\begin{eqnarray}
\bar{S}^{(1)}(k) &=& e B \Bigg\{
\int_{0}^{\infty} s ds\, e^{i s [(k_0+\mu)^2-m^2-\mathbf{k}^2+i\epsilon]} 
+2i\pi  \theta(|\mu|-|k_0|) \theta(-k_0\mu) 
\delta^{\prime}\left[(k_0+\mu)^2-m^2-\mathbf{k}^2 \right]
\Bigg\}\nonumber\\
&&\times \left[(k_0+\mu)\gamma^{0}-k^3\gamma^3+m\right]\gamma^1\gamma^2,
\label{S-prop-time-mu3}
\end{eqnarray}
where $\bar{S}^{(1)}(k)$ splits naturally into the ``vacuum" and ``matter" parts, with the latter 
containing the $\delta$ function. [Note that the vacuum part is not precisely reflecting the 
nature of the first contribution, because it depends on the chemical potential.] It is 
convenient to treat the two pieces separately in the calculation of the self-energy.

To linear order in the magnetic field, the vacuum part of the self-energy is given by
\begin{eqnarray}
\bar{\Sigma}_{\rm vac}^{(1)} (p) &=& 8 i\pi\alpha eB  \int_{0}^{\infty} d\tau \int_{0}^{\infty} s ds\, 
\int\frac{ d^4 k}{(2\pi)^4} e^{i s \left[(k_0+\mu)^2-m^2-\mathbf{k}^2\right]+i\tau \left[(p_0-k_0)^2-
(\mathbf{p}-\mathbf{k})^2\right]}
\left[(k_0+\mu)\gamma^{0}-k^3\gamma^3\right]\gamma^1\gamma^2\nonumber\\
&=& - \frac{\alpha eB}{2\pi} \left[(p_0+\mu)\gamma^{0}-p^3\gamma^3\right]\gamma^1\gamma^2 
\int_{0}^{\infty} \int_{0}^{\infty} \frac{s \tau ds d\tau}{(s+\tau)^3}
e^{-ism^2+i\frac{s\tau}{s+\tau}\left[(p_0+\mu)^2-\mathbf{p}^2 \right]} 
\nonumber\\
&=& - \frac{\alpha eB}{2\pi}\,i\gamma^1\gamma^2 \frac{(p_0+\mu)\gamma^{0}-p^3\gamma^3}
{(p_0+\mu)^2-\mathbf{p}^2}
\left[1+\frac{m^2}{(p_0+\mu)^2-\mathbf{p}^2}\left(\ln\frac{|m^2+\mathbf{p}^2-(p_0+\mu)^2|}{m^2}-i
\pi \theta[...] \right) \right],
\label{self-energy-vacuum}
\end{eqnarray}
where the imaginary part is nonzero when $(p_0+\mu)^2-\mathbf{p}^2>m^2$. Note that this 
expression simplifies a lot in the chiral limit.

To the same linear order in the magnetic field, the matter part of the self-energy is given by 
\begin{equation}
\bar{\Sigma}_{\rm mat}^{(1)} (p) = -\frac{\alpha eB}{\pi^2}\,i\gamma^1\gamma^2
\int d^4 k \frac{(k_0+\mu)\gamma^{0}-k^3\gamma^3}{(p_0-k_0)^2-(\mathbf{p}-\mathbf{k})^2}\, 
\theta(|\mu|-|k_0|) \theta(-k_0\mu) 
\delta^{\prime}\left[(k_0+\mu)^2-m^2-\mathbf{k}^2 \right].
\label{self-energy-matter}
\end{equation}
We would like to emphasize that Eqs.~(\ref{self-energy-vacuum}) and (\ref{self-energy-matter}) imply 
that only the chiral asymmetric structures $\Delta$ and $\mu_5$ are generated in the fermion self-energy 
in the linear in $B$ approximation. Therefore, we have the following contributions to the chiral shift and 
chiral chemical potential terms in Eq.~(\ref{self-energy-general}):
\begin{eqnarray}
\Delta_{\rm vac}(p)&=&-\frac{i \alpha (p_0+\mu) eB }{2\pi \left[(p_0+\mu)^2 -\mathbf{p}^2\right]
^2}\Bigg[ 
(p_0+\mu)^2 -\mathbf{p}^2
+m^2\ln\frac{|m^2+\mathbf{p}^2-(p_0+\mu)^2 |}{m^2}-i\pi \, m^2 \, \theta \left[(p_0+\mu)^2 -
\mathbf{p}^2-m^2\right]
\Bigg], 
\label{Delta-vac}\\
\mu^{\rm vac}_5(p) &=&-p^3\Delta_{\rm vac}/(p_0+\mu),
\label{parameters-vacuum}
\end{eqnarray}
\begin{eqnarray}
\Delta_{\rm mat}(p)&=&
-\frac{\alpha e B }{4\pi|\mathbf{p}|}\ln\frac{p_0^2-(|\mathbf{p}|-\sqrt{\mu^2-m^2})^2}{p_0^2-(|\mathbf{p}|+\sqrt{\mu^2-m^2})^2}
+ \frac{\alpha e B \sqrt{\mu^2-m^2}}{2\pi [\mathbf{p}^2-(p_0+\mu)^2]}
+\frac{\alpha e B m^2 (p_0+\mu)}{2\pi \left[(p_0+\mu)^2-\mathbf{p}^2\right]^2}
\ln\frac{\mu-\sqrt{\mu^2-m^2}}{\mu+\sqrt{\mu^2-m^2}}\nonumber\\
&+&\frac{\alpha e B }{8\pi|\mathbf{p}|}\left(\frac{(p_0+\mu+|\mathbf{p}|)^2-m^2}{(p_0+\mu+|\mathbf{p}|)^2}
\ln\frac{p_0+|\mathbf{p}|-\sqrt{\mu^2-m^2}}{p_0+|\mathbf{p}|+\sqrt{\mu^2-m^2}}
-\frac{(p_0+\mu-|\mathbf{p}|)^2-m^2}{(p_0+\mu-|\mathbf{p}|)^2}
\ln\frac{p_0-|\mathbf{p}|-\sqrt{\mu^2-m^2}}{p_0-|\mathbf{p}|+\sqrt{\mu^2-m^2}}\right),
\label{Delta-mat}
\end{eqnarray}
\begin{eqnarray}
\mu^{\rm mat}_5(p)&=&-\frac{\alpha eBp^3 }{2\pi\mathbf{p}^2}\Bigg[ 
\frac{(p_0+\mu)\sqrt{\mu^2-m^2}}{\mathbf{p}^2-(p_0+\mu)^2} 
+\frac{\mu^2+p_0(p_0+\mu)-\mathbf{p}^2+m^2}{2p\mu } 
\ln\frac{p_0^2-(|\mathbf{p}|-\sqrt{\mu^2-m^2})^2}{p_0^2-(|\mathbf{p}|+\sqrt{\mu^2-m^2})^2}
\nonumber\\
&+& 
\frac{3(p_0+\mu)(p_0+\mu+|\mathbf{p}|)^2+m^2(p_0+\mu+2|\mathbf{p}|)}{4|\mathbf{p}|(p_0+\mu+|\mathbf{p}|)^2}
\ln\frac{p_0+|\mathbf{p}|-\sqrt{\mu^2-m^2}}{p_0+|\mathbf{p}|+\sqrt{\mu^2-m^2}}\nonumber\\
&-&\frac{3(p_0+\mu)(p_0+\mu-|\mathbf{p}|)^2+m^2(p_0+\mu-2|\mathbf{p}|)}{4|\mathbf{p}|(p_0+\mu-|\mathbf{p}|)^2}
\ln\frac{p_0-|\mathbf{p}|-\sqrt{\mu^2-m^2}}{p_0-|\mathbf{p}|+\sqrt{\mu^2-m^2}}
\Bigg].
\label{parameters-matter}
\end{eqnarray}
Equation~(\ref{parameters-matter}) shows that, as mentioned above, $\mu_5(p)$ is indeed an odd function of $p^3$, and, therefore, 
it does not break parity.

It is useful to consider some particular limits of the obtained expressions. The first interesting 
case corresponds to the behavior of the chiral shift and chiral chemical potential on the Fermi 
surface, i.e., for $p_0\to 0$ and $|\mathbf{p}|\to p_F\equiv \sqrt{\mu^2-m^2}$. We have
\begin{eqnarray}
\Delta=\Delta_{\rm mat}+\Delta_{\rm vac} &\simeq & \frac{\alpha e B \mu }{\pi m^2}
\left(\ln\frac{m^2}{2\mu\left(|\mathbf{p}|-p_F\right)}-1\right),\\
\mu_5=\mu^{\rm mat}_5+\mu^{\rm vac}_5 &\simeq &  -\frac{\alpha e B \mu \cos\theta}{\pi m^2}
\left(\ln\frac{m^2}{2\mu\left(|\mathbf{p}|-p_F\right)}-1\right),
\end{eqnarray}
where $\cos\theta=p^3/p$; i.e., $\theta$ is the angle between the magnetic field and momentum. 
Furthermore, Eqs.~(\ref{Delta-vac}) through
(\ref{parameters-matter}) simplify strongly in the chiral limit, where, for $\mu>0$,
\begin{eqnarray}
\bar{\Sigma}^{(1)} (p) &\simeq & 
-\frac{\alpha e B }{2\pi}\gamma^3\gamma^5\left[\frac{p_0+2\mu}{(p_0+\mu)^2-\mathbf{p}^2}
-\frac{1}{4|\mathbf{p}|}\ln\frac{p_0^2-(|\mathbf{p}|+\mu)^2}{p_0^2-(|\mathbf{p}|-\mu)^2}\right]
\nonumber\\
&+&\frac{\alpha e B }{2\pi}\gamma^0\gamma^5\frac{p^3}{\mathbf{p}^2}\left[
\frac{\mu(p_0+\mu)+\mathbf{p}^2}{(p_0+\mu)^2-\mathbf{p}^2}
+\frac{\mu-p_0}{4 |\mathbf{p}|}\ln\frac{p_0^2-(|\mathbf{p}|-\mu)^2}{p_0^2-(|\mathbf{p}|+\mu)^2}
\right].
\end{eqnarray}

\section{Chiral asymmetry}
\label{Sec:Asymmetry}

The dispersion relations for fermion quasiparticles in a weak magnetic field can be formally 
obtained by considering the location of the poles of the fermion propagator. As we discussed in Sec.~\ref{Sec:Model}, in the limit 
of large pseudomomentum or weak magnetic field (i.e., $\mathbf{k}^2_{\perp} \gg |eB|$), the effects 
of the Schwinger phase can be neglected and pseudomomentum can be interpreted as 
an approximate (or ``quasiclassical") fermion's momentum. Then the poles of the fermion propagator
are defined by the following equation:
\begin{equation}
\mbox{det}\left[i\bar{S}^{-1}(p)-\Sigma(p)\right] =0.
\label{det=0}
\end{equation}

To determine the dispersion relations from Eq.~(\ref{det=0}), we should define the inverse free propagator 
in the pseudomomentum representation. This is not difficult to do by following the procedure given in 
Sec.~\ref{Sec:Landau-level-expansion}. The inverse free propagator in the coordinate space is defined as follows:
\begin{eqnarray}
iS^{-1}(x,y) &=& \left(i\gamma^{\nu}{\cal D}_{\nu}+\mu\gamma^0-m\right)\delta^4(x-y).
\label{inverse-S-coordinate-space}
\end{eqnarray}
By making use of Eq.~(\ref{completeness}), one can rewrite the inverse free propagator (\ref{inverse-S-coordinate-space}) 
in the form
\begin{equation}
iS^{-1}(x,y) =\sum_{N=0}^{\infty} \int\frac{dp_0 dp^3 dp\,\,e^{-ip_0(x_0-y_0)+ip^3(x^3-y^3)}}{(2\pi)^2}
\left[(p_0+\mu)\gamma^0-p^3\gamma^3
-(\bm{\pi}_{\perp}\cdot\bm{\gamma}_{\perp})-m
\right] \psi_{Np}(\mathbf{r}_{\perp}) \psi_{Np}^{*}(\mathbf{r}_{\perp}^\prime).
\end{equation}
After taking into account the identity in Eq.~(\ref{pi-gamma-relation}) and the table 
integral in Eq.~(\ref{HnHm-integral}), we can easily perform the integration over the quantum 
number $p$. Just like in the case of the self-energy, the result takes the form of a 
product of the standard Schwinger phase and a translationally invariant function, i.e.,
\begin{equation}
iS^{-1}(x,y) = e^{i\Phi(x,y)} i\bar{S}^{-1}(x-y) .
\end{equation}
The translationally invariant function is given by
\begin{eqnarray}
i\bar{S}^{-1}(x) &=&\frac{e^{-\xi/2}}{2\pi\ell^2}
\sum_{n=0}^{\infty}\int \frac{dp_0 dp^3 }{(2\pi)^2}  e^{-ip_0x_0+ip^3x^3} \Big\{
\left[(p_0+\mu)\gamma^0 -p^3\gamma^3-m\right] \left[ L_{n}(\xi){\cal P}_{-}+L_{n-1}(\xi){\cal P}_{+}\right]
\nonumber\\
&&+\frac{i}{\ell^2}(\mathbf{r}_{\perp}\cdot\bm{\gamma}_{\perp}) L^{1}_{n-1}(\xi)
\Big\} ,
\end{eqnarray}
where $\xi = \mathbf{r}_{\perp}^2/(2\ell^2)$. By performing the Fourier transform, we finally arrive at 
the following expansion of the translation invariant part of the inverse free propagator over Landau levels 
[compare with the corresponding expansion of the self-energy in Eq.~(\ref{self-energy-LL})]:
\begin{equation}
i \bar{S}^{-1}(p) = 2  e^{-p_{\perp}^2 \ell^2}  \sum\limits_{n=0}^{\infty} (-1)^n\Big\{
\left[(p_0+\mu)\gamma^0 - p^3\gamma^3 -m\right] 
\left[{\cal P}_{-} L_n(2p_{\perp}^2 \ell^2)-{\cal P}_{+} L_{n-1}(2p_{\perp}^2 \ell^2) \right] 
+2 (\bm{\gamma}_\perp\cdot \bm{p}_{\perp})
L^1_{n-1}(2p_{\perp}^2 \ell^2)\Big\} .
\label{inverse-bare-propagator}
\end{equation}
Interestingly, by performing the summation over Landau levels in this expression using the following well-known formula 
\cite{Gradstein_Ryzhik}:  
\begin{equation}
\sum\limits_{n=0}^{\infty} z^n L_n^\alpha(x) = \frac{1}{(1-z)^{1+\alpha}}\exp\left(\frac{x z}{z-1}\right),
\label{table-sum}
\end{equation}
we obtain
\begin{equation}
i\bar{S}^{-1}(p) =(p_0 +\mu)\gamma^0 -(\bm{\gamma}_\perp\cdot \mathbf{p}_{\perp})- p^3\gamma^3 -m .
\label{inverse-free-momentum}
\end{equation}
This is a remarkable result, because it means that the translation invariant part of the inverse free propagator in 
a magnetic field is identical to the inverse free propagator in the absence of a magnetic field. Therefore, for the 
inverse free propagator, only the Schwinger phase contains information about the presence of a magnetic field.
 
For the free propagator in the weak field limit, the dependence on the Landau level index [which is 
the eigenvalue of the operator $-\frac{1}{2}(\bm{\pi}_\perp\cdot\bm{\gamma}_\perp)^2\ell^2$] can be 
unambiguously replaced by the square of the transverse momentum, i.e., $2n|eB| \to \mathbf{p}_\perp^2$. 
Therefore, when using the pseudomomentum representation in Eq.~(\ref{det=0}), we can interpret 
$\mathbf{p}_\perp^2$ as a convenient shorthand substitution for $2n|eB|$. Indeed, this is natural in 
the weak field limit, when the quantization of Landau levels is largely irrelevant. This implies the 
standard dispersion relation $p_0=-\mu \pm \sqrt{\mathbf{p}_{\perp}^2+p_3^2+m^2}$, or equivalently 
$p_0=-\mu \pm \sqrt{2n|eB|+p_3^2+m^2}$ after the substitution $\mathbf{p}_\perp^2\to 2n|eB|$.

\begin{figure}[t]
 \includegraphics[width=0.45\textwidth]{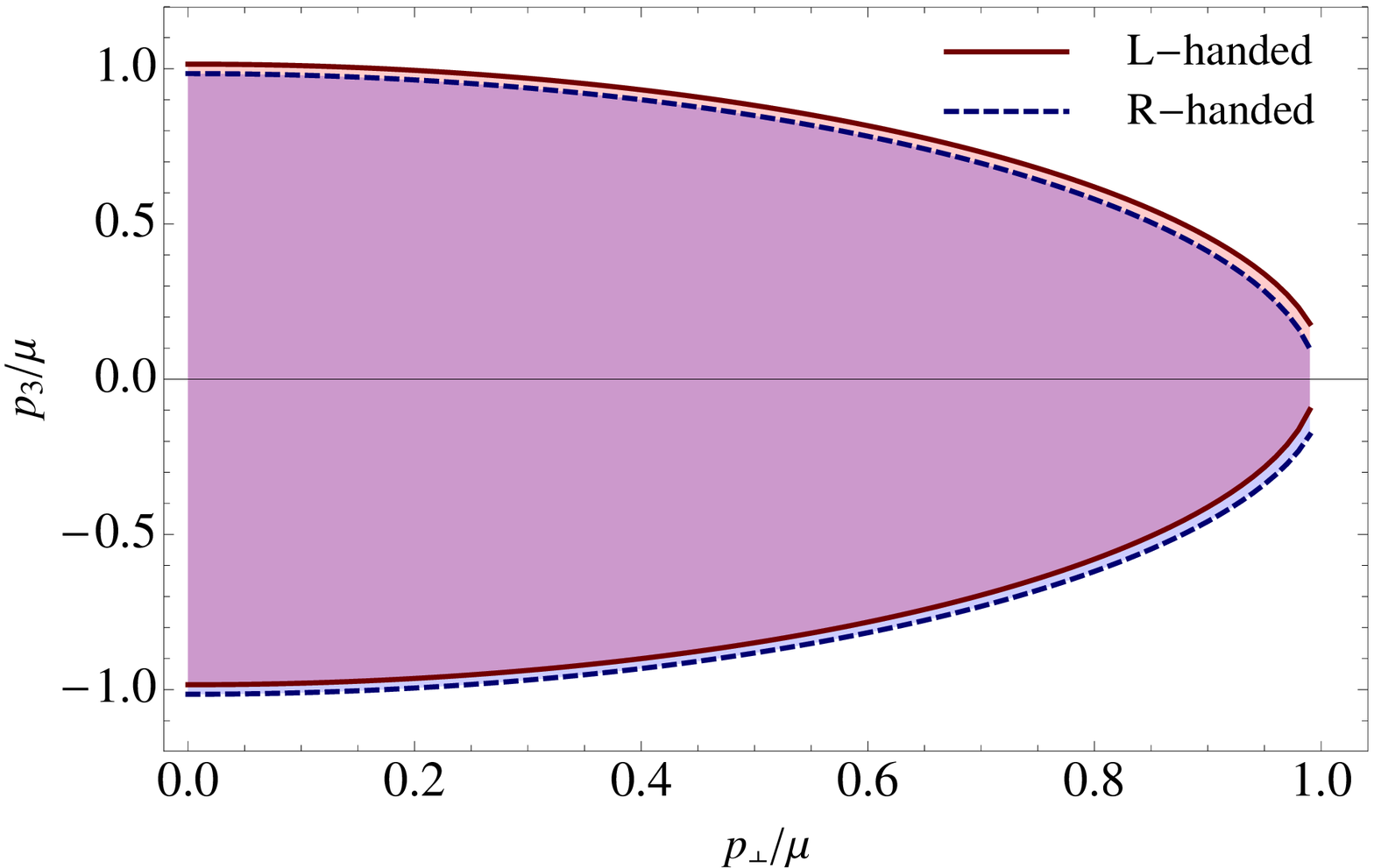}\hspace{0.05\textwidth}
\includegraphics[width=0.45\textwidth]{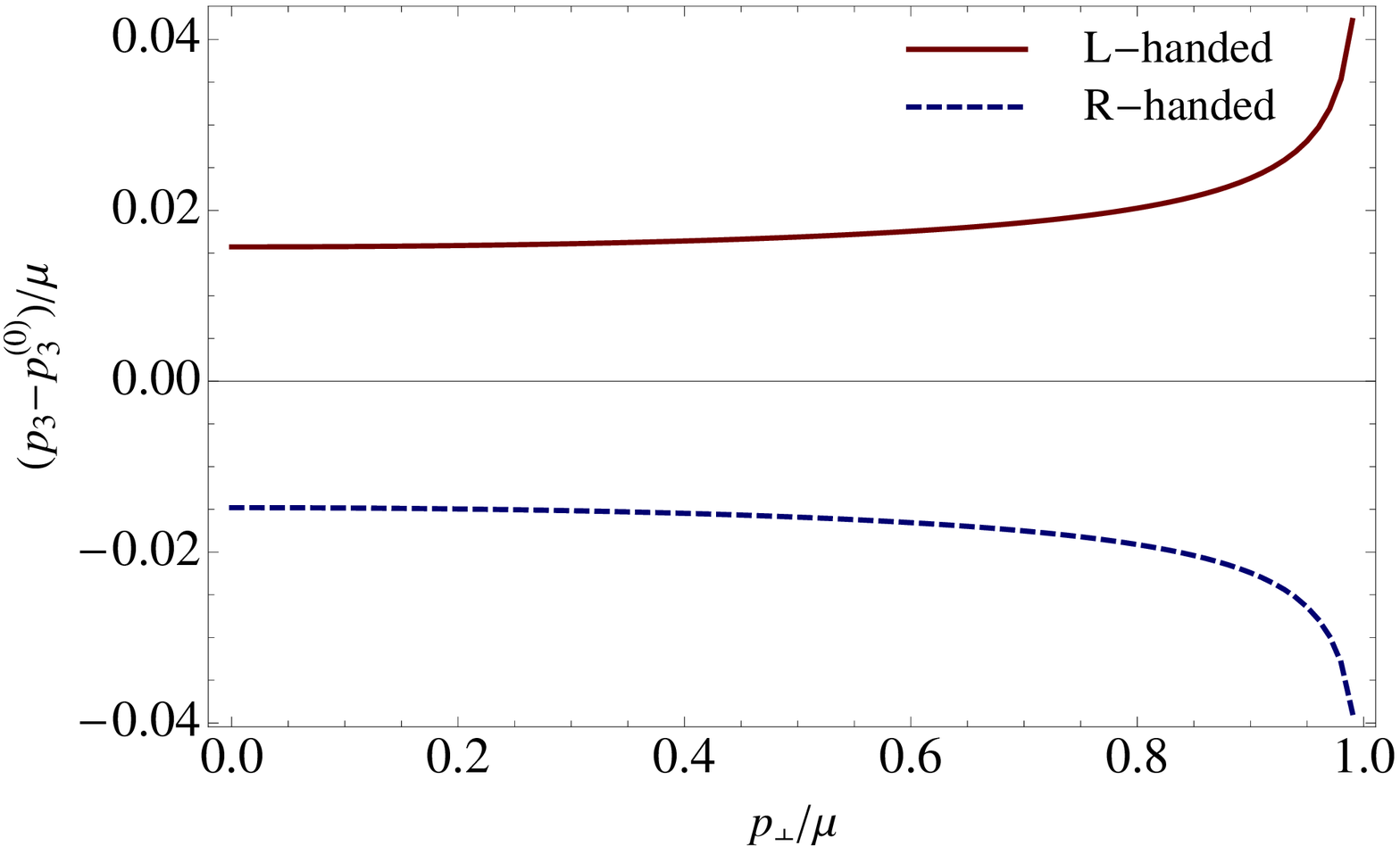}
\caption{Asymmetry of the Fermi surface for predominantly left-handed and right-handed particles 
for $|eB|=0.1\mu^2$ and $\alpha=1/137$.}
\label{fig-fermi-surf}
\end{figure}

By making use of the chiral representation of the Dirac $\gamma$ matrices, the inverse free propagator 
(\ref{inverse-free-momentum}), and the self-energy in the weak magnetic field limit, Eq.~(\ref{det=0}) 
can be rewritten in the following equivalent form:
\begin{equation}
\mbox{det}\left(
\begin{array}{cc}
p_0 +\mu -(\bm{\sigma}_\perp\cdot \mathbf{p}_{\perp})+(\Delta- p^3)\sigma^3 +\mu_5& m \\
m & p_0 +\mu +(\bm{\sigma}_\perp\cdot \mathbf{p}_{\perp})+ (\Delta+ p^3)\sigma^3-\mu_5
\end{array}
\right)=0,
\label{det=0-more1}
\end{equation}
where $\bm{\sigma}$ are Pauli matrices. Calculating the determinant, we obtain
\begin{equation}
\left[(p_0 +\mu-\mu_5)^2-\mathbf{p}_{\perp}^2-(p^3+\Delta)^2\right]
\left[(p_0 +\mu+\mu_5)^2-\mathbf{p}_{\perp}^2-(p^3-\Delta)^2\right]
-2m^2\left[(p_0 +\mu)^2+\Delta^2-\mathbf{p}_{\perp}^2-p_3^2-\mu_5^2\right]+m^4=0.
\label{det=0-more2}
\end{equation}
This expression can be factorized to produce two equations for predominantly left-handed 
and predominantly right-handed particles:
\begin{eqnarray}
(p_0 +\mu)^2-\mathbf{p}_{\perp}^2-p_3^2-m^2-\Delta^2+\mu_5^2
-2\sqrt{(p^3\Delta +\mu_5(p_0 +\mu))^2+m^2(\Delta^2-\mu_5^2)}
=0,
\label{det=0-more3}\\
(p_0 +\mu)^2-\mathbf{p}_{\perp}^2-p_3^2-m^2-\Delta^2+\mu_5^2
+2\sqrt{(p^3\Delta +\mu_5(p_0 +\mu))^2+m^2(\Delta^2-\mu_5^2)}
=0.
\label{det=0-more4}
\end{eqnarray}
By making use of the analytical results for the self-energy obtained in the previous section [see Eqs.~(\ref{Delta-vac}) --
(\ref{parameters-matter}) and the dispersion relations that follow from Eqs.~(\ref{det=0-more3}) and (\ref{det=0-more4})],
we can easily write down the equations for the Fermi surfaces of both types of particles. Namely, we take $p^0=0$ and 
solve for $p^3$ as a function of $p_\perp$. The results are shown in the left panel of Fig.~\ref{fig-fermi-surf} in the
case of the physical value of the fine structure constant ($\alpha=1/137$) and the magnetic field $|eB|=0.1\mu^2$. 
In order to clearly demonstrate the magnitude of the effect, in the right panel of Fig.~\ref{fig-fermi-surf} we 
also plot the difference between the longitudinal momenta with and without the inclusion of the interaction 
induced chiral asymmetry.

%%%%%%%%%%%%%%%%%%%%%%%%%%%%%%%%%%%%%%%%%%%%%%%%%%%%%
%%%%%%%%%%%%%%%%%%%%%%%%%%%%%%%%%%%%%%%%%%%%%%%%%%%%%

As the results in Fig.~\ref{fig-fermi-surf} demonstrate, the Fermi surface of the predominantly 
left-handed particles is slightly shifted in the direction of the magnetic field, while the Fermi surface of 
the predominantly right-handed particles is slightly shifted in the direction opposite of the magnetic field. 
This is in qualitative agreement with the finding in the NJL model \cite{chiral-shift-2}. In the case of QED
with its long-range interaction, however, the chiral asymmetry of the Fermi surfaces comes not only from 
the $\Delta$ function, but also from the new function $\mu_5(p)\equiv p_3 f(p)$. Also, unlike in the NJL 
model, both of these functions have a nontrivial dependence on the particles' momenta. In particular, they
reveal a logarithmic enhancement of the asymmetry near the Fermi surface.

\section{Discussion and Conclusion}
\label{Sec:Conclusion}

Studying the fermion self-energy in dense QED in a magnetic field, we confirm, as suggested 
by the corresponding studies in the NJL model \cite{chiral-shift-1,chiral-shift-2}, that nonzero radiative 
corrections to the axial current found in Ref.~\cite{Gorbar:2013upa} are connected with the presence of 
chiral asymmetry in higher Landau levels induced by interaction. Our result for the fermion self-energy, 
obtained perturbatively in the coupling constant and in linear order 
in the external magnetic field, reveals the presence of two chirally asymmetric structures. One of them 
is the chiral shift function, analogous to the one previously obtained in the NJL model. The other one 
is a new structure that resembles the chiral chemical potential. However, unlike the chiral chemical 
potential, it preserves parity because it is an odd function of the momentum
directed along the magnetic field. Note that this dependence on momentum is dictated by the 
parity symmetry. Since QED in a magnetic field is invariant under parity and the self-energy is 
obtained in a perturbation theory, parity cannot be broken. The term $\mu_5\bar{\psi}\gamma^0\gamma^5\psi$ 
is not parity invariant unless $\mu_5$ is an odd function of momentum $\pi^3=-i\partial_3$. The same 
argument also ensures that there is no electric current along the direction of the magnetic field, which 
would be present due to the chiral magnetic effect if one had $\mu_5=\mbox{const}$.

The current study of the chiral asymmetry in the ground state of a cold dense QED plasma is in
qualitative agreement with the earlier study of the asymmetry in the NJL model \cite{chiral-shift-1,chiral-shift-2}.
In particular, the Fermi surfaces of the left- and right-handed fermions are shifted relative to each 
other in momentum space in the direction of the magnetic field. The value of the shift appears to be rather 
large and, thus, may have important phenomenological implications. For example, 
a relativistic matter in stars, in which the chiral shift parameter is nonvanishing, will cause neutrinos to scatter off 
asymmetrically. As proposed in Refs. \cite{chiral-shift-1,chiral-shift-2}, this
can provide a new mechanism for the pulsar kicks. Indeed, when the trapped neutrinos in a protoneutron star
experience multiple elastic scattering on the chirally asymmetric state, they build up an anisotropic momentum
distribution and, thus, provide a kick after leaving the star.  

As should be clear from the physical meaning of the results obtained in this study, the chiral asymmetry
in the ground state of the cold dense QED plasma is the main source of the radiative corrections to the
axial current density calculated in Ref.~\cite{Gorbar:2013upa}. Indeed, as the direct calculations show,
the corresponding corrections originate from the perturbative self-energy contribution in the expansion 
of the fermion propagator. This observation is also consistent with the fact that the result for the axial current
is particularly sensitive to the fermion states in the vicinity of the Fermi surface, i.e., the region of the phase 
space where the chiral asymmetry is most important.

In the future, it will be interesting to generalize the current study to the case of strong magnetic fields utilizing 
the expansion over Landau levels and using numerical methods in calculations. In addition, the role of the 
photon screening effects should be understood in detail. It will be also interesting to clarify the connection 
between the chiral shift in relativistic systems in a magnetic field and a class of condensed matter systems 
with Weyl quasiparticles, which reveal similar properties even in the absence of external fields \cite{Wan,Balents}.

\begin{acknowledgments}
The work of E.V.G. was supported partially by the European FP7 program, Grant No. SIMTECH 246937, 
and SFFR of Ukraine, Grant No.~F53.2/028. The work of V.A.M. was supported by the Natural Sciences 
and Engineering Research Council of Canada and, partially, by the National Science Foundation under 
Grant No. PHYS-1066293 and the hospitality of the Aspen Center for Physics. He is also grateful to the 
Galileo Galilei Institute for Theoretical Physics and the INFN for their hospitality and partial support during 
the completion of this work. The work of I.A.S. and X.W. was supported, in part, by the U.S. National 
Science Foundation under Grant No.~PHY-0969844.
\end{acknowledgments}

\end{document}